\documentclass[twocolumn,showpacs,a4paper,nofootinbib,tightenlines,floats]{revtex4}
\usepackage{bm}
\usepackage{latexsym}
\usepackage{dcolumn}
\usepackage{color}
\usepackage{amsfonts,amssymb}
\usepackage{graphicx,epsfig}
\usepackage{psfrag}
\usepackage{amsmath,amssymb}
\usepackage{graphicx,epsfig}
\begin{document}
\newcommand {\be}{\begin{equation}}
 \newcommand {\ee}{\end{equation}}
 \newcommand {\bea}{\begin{array}}
 \newcommand {\cl}{\centerline}
 \newcommand {\eea}{\end{array}}
 \renewcommand {\theequation}{\thesection.\arabic{equation}}
 \newcommand{\red}[1]{\textcolor[rgb]{1.00,0.00,0.00}{#1}}%
 \newcommand {\newsection}{\setcounter{equation}{0}\section}

\title{\textbf{Simplifying the algebra of first class constraints,\\ examples on $SO(3)$ and $SO(4)$}}

\author{M. Dehghani\footnote{mdehghani@ph.iut.ac.ir} \\  A. Shirzad\footnote{shirzad@ipm.ir}}
\affiliation{Department of Physics, Isfahan University of Technology \\
P.O.Box 84156-83111, Isfahan, IRAN, \\
School of Physics, Institute for Research in Fundamental Sciences (IPM)\\
P.O.Box 19395-5531, Tehran, IRAN.}

\date{\today}

\pacs{}

\begin{abstract}
We discuss the problem of non abelian constrained systems and the
origin of appearance of non abelian algebras. We show that it is
possible, in principle, to change a non abelian system to an
abelian one, at least locally. Our method is based on solutions of
the differential equations due to the algebra of first class
constraints. We suggest that multiplicative constraints should be
regularized at each step of calculation. Some examples on $SO(3)$
and $SO(4)$ algebras are discussed. We show that they can be
replaced by abelian constraints.

\textbf{Keywords}: First class system, Constraint algebra,
Abelianization.

\end{abstract}

\maketitle
\section{Introduction \label{sec1}}
The general formalism of constrained systems seems sometimes very
complicated. The reason is the complicated and extensive algebra
of constraints. Consider a first class system given by constraints
$\phi_a$ and the canonical Hamiltonian $H_c$. The most general
form of the algebra of Poisson brackets is \cite{HENOUX}:
 \be \begin{array}{c}
       \{\phi_a,\phi_b\}=C_{ab}^c\phi_c \\
       \{\phi_a,H_c\}=V_{a}^b\phi_b \\
     \end{array} \label{intro1}\ee
The coefficients $C_{ab}^c$ and $V_a^b$ are called structure
functions. In general these coefficients may be functions of phase
space variables $(q,p)$. If one wishes to keep the track of levels
of consistency, additional labels showing the level should be
assigned to the constraints and the algebra would be much more
complicated. This complication causes so many difficulties in
proving general statements in the context of constrained systems
and makes this field of study difficult to follow. However, in the
real models of gauge systems we never encounter a problem in which
the complete set of structure functions $C_{ab}^c$ and $V_a^b$ are
present or depend on $(q,p)$. The simplest possibility is
$C_{ab}^c=0$ and $V_{a}^{b}=\textrm{cons.}$. Such systems are
called abelian first class constrained systems. As we will see in
the next section this is the case for quadratic Lagrangians (with
respect to velocities and coordinates) which include most physical
models. A non abelian system requires a Lagrangian or Hamiltonian
with higher powers, or more complicated functional dependence on
the corresponding variables.

On the other hand, it is well known \cite{pons} that the choice of
constraints in not unique. In other words, different sets of
constraints may describe the same constraint surface. We say that
the two sets of constraints $\phi_a(q,p)$, $a=1,\cdots m$ and
$\phi_{a'}(q,p)$, $a'=1,\cdots m'$ are equivalent if
$\phi_a(q,p)=0\Leftrightarrow\phi_{a'}(q,p)=0$. Then it is obvious
that in general, one can write:
  \be \phi_a(q,p)=\sum_{a'=1}^{m'}M_{aa'}\phi_{a'}(q,p) \;\;\;\;\;\;\ a=1,\cdots  m.\label{mmatrix}\ee
A system is called reducible if it can be converted to one with
less number of constraints. Transforming from the set
$\phi_a(q,p)$ to the set $\phi_{a'}(q,p)$ is called a
"redefinition" of constraints.

We wish to use the redefinition process to make the algebra of the
constraints as simple as possible. The best situation is one with
abelian algebra. Then the system is said to be abelianized. We can
also make use of the "canonical transformations" (CT's ) whenever
necessary. The first tool, i.e. redefinition of the constraints,
changes the variables describing the constraint surface only,
while the second one, i.e. CT, change the coordinates of the whole
phase space canonically. It should be emphasized that the Poisson
brackets are invariant under CT, while the redefinition procedure
may lead to a different algebra of Poisson brackets.

Converting a system of first class constraints to an abelian
system, if it is possible, has so many advantages. In fact in any
analysis based on the constraint structure of a gauge system one
encounters the complicated algebra of Poisson brackets of first
class constraints and one can not show the essential features of
the problem in clear formulas. For example a closed formula for
the generating function of gauge transformations in terms of the
constraints, is not given so far. There have been proposed only
some rules and instructions in this regard for the most general
case of structure functions with arbitrary dependence on $(q,p)$
\cite{HENOUX,HENOUX2}.

There is a similar situation for the BRST charge which is needed
to quantize a gauge system. In fact, this charge can be written as
an infinite expansion in terms of recursive Poisson brackets of
structure functions which in general make an open algebra which
does not terminate. For an abelian system, on the other hand,
besides clarity and simplicity in understanding the structure of
physical as well as gauged degrees of freedom, one is able to
write down the generator of gauge transformations in a closed form
\cite{shirzadshabani} and the expansion of the BRST charge
terminates after the first term \cite{HENOUXREPORT}.

As we will see through the paper, non abelian constraints may stem
from the choice of coordinates for description of constraint
surface. In this view, the abelianization procedure is a
redefinition and purgation process for the constraints which make
it possible to describe the constraint surface with a set of
suitable (i.e. commuting ) coordinates.

Our aim in this paper is to show that some first class constrained
systems can in principle be abelianized, at least locally. In
fact, one can say that the origin of non abelianity stem from bad
choice of the canonical coordinates of the whole phase space as
well as the variables describing the constraint surface.
Therefore, the abelianization means that one tries to find the
suitable coordinates in which the algebra of constraints is
abelian.

Local abelianization has been shown, in principle, to be possible
\cite{HENOUX,gogi} through solving constraint equations,
$\phi_a=0$ to find a number of coordinates $\xi_a$ in terms of
other variables $\widetilde{\xi}$ as:
   \be \xi_{a'}=f_{a'}(\xi) \ee
Then the new constraints $\psi_{a'}=\xi_{a'}-f_{a'}(\xi)$ have
Poisson brackets which are independent of $\xi_{a'}$. Hence, the
only way that the Poisson brackets may vanish on the constraint
surface $\psi_{a'}=0$ is that they vanish strongly. In \cite{L1}
it is argued that if one maps each constraint to the surface of
other constraints they would be abelianized. However, it is
asserted in \cite{L3} that for gauge systems such as $SO(3)$ the
maximality conditions is violated, hence the sufficient condition
for $SO(3)$  to be abelianizable is not satisfied. Another method
is proposed for abelianization \cite{gogi} which is based on
finding a complicated solution for the matrix $M$ in
(\ref{mmatrix}) such that the new set of constraints are abelian.

In this paper we try to study the problem of abelianization in a
systematic way from the point of view of differential equations
coming from the algebra of constraints. Our method is based on
finding suitable coordinates to describe the constraint surface.
We do this by solving differential equations due to Poisson
brackets of constraints with one momentum constraint. We will find
that in this way one would naturally lead to simple coordinates of
the constraint surface. This method will be discussed in details
in section \ref{sec3}. Before that we will discuss in the next
section some general aspects of abelian and non abelian nature of
first class systems. We will also discuss the problem of
regularity of the constrains. We suggest that the multiplicative
constraints should be regularized before abelianization. Section
\ref{so13} denotes detailed calculations concerning abelianization
of a first class system with $SO(3)$ algebra. This important
example shows the general features of the abelianization
procedure. In section \ref{conclusion} we give our conclusions.
\section{How non abelian constraints may happen? \label{sec2}}
Let us first consider a simple example to see in what sense a non
abelian system of first class constraints may emerge. Suppose, in
a system with $q_1$ and $q_2$ as coordinates, we are given two
first class constraints
     \be  \begin{array}{cc}
     \phi_1=p_1e^{\alpha q_2} & \phi_2=p_2e^{-\beta q_1} \\
     \end{array} \label{eq1}\ee
where $\alpha $ and $\beta$ are constants. Clearly we have
     \be \{\phi_1,\phi_2\}=(\alpha e^{-\beta q_1})\phi_1
     +(\beta e^{\alpha q_2})\phi_2 \label{eq1e1}\ee
which exhibits the non abelian feature of the system. It is,
however, obvious that the constraints surface $\phi_1=0$ and
$\phi_2=0$ is equivalent to the surface described by $p_1=0$ and
$p_2$, since the exponentials does not vanish in the finite range
of their arguments. Clearly, the constraints $p_1$ and $p_2$ are
abelian. In general the situation is not so obvious, and the
algebraic structure of the Poisson brackets may be so complicated
that one is not able to recognize the best and simplest phase
space coordinates describing the constraints surface.

It is also possible to inspect some features of non abelian
systems by power counting. Suppose $\phi_i(z^m)$ is any constraint
which can be written as a polynomial of order $m$ with respect to
phase space coordinates $z_{\mu}, \mu=1,\cdots 2N$. For a first
class system the algebra of Poisson brackets reads
     \be \{\phi_i(z^m),\phi_j(z^n)\}=a_{ij}^k(z^r)\phi_k(z^s).
  \label{eqadded1}\ee
Since Poisson bracket requires two times of differentiation, we
should have
     \be m+n-2=r+s. \label{eqadded2}\ee
If the constraints are linear with respect to $z_{\mu}$, i.e.
$m=n=1$, then a first class system can be achieved just for
$a_{ij}^k=0$. In other words linear first class constraints are
essentially abelian. For quadratic Lagrangians, which is the case
for most physical systems, the primary constraints which emerge
due to singularity of Hessian (the matrix of second derivatives
with respect to velocities), are necessarily linear. Since the
Hamiltonian is also quadratic, the secondary constraints at any
level would be linear, too. So, for the wide class of gauge
systems with quadratic Lagrangians the system is abelian by
itself. The above analysis shows that the problem of
abelianization maybe converted to the problem of linearization. In
other words if we are able to abelianize a first class system this
means that there can be found suitable coordinates in which the
constraint surface is described by constant values of some phase
space coordinates. Specially we can choose a basis in which the
constraints are some momenta.

For example a constrained system given by $\phi_1=xp_x+yp_y$,
$\phi_2=\frac{1}{2}p_x^2$ and $\phi_3=\frac{1}{2}p_y^2$ obeys the
non abelian algebra $\{\phi_1,\phi_2\}=\phi_2$ and
$\{\phi_1,\phi_3\}=\phi_3$. This system reduces to the constraints
$p_x$ and $p_y$, which are linear as well as abelian.

Coming back to the Eq's. (\ref{eqadded1}) and (\ref{eqadded2}),
assume a system in which the constraints are quadratic homogenous
functions of $z_{\mu}$ which leads to $r=0$. In other words, for
quadratic constraints we may have a non abelian closed algebra of
Poisson brackets only with constant structure functions
$a_{ij}^k$. For example in a system with $x$, $y$ and $z$ as
coordinates, the quadratic constraints $\phi_1=yp_z-zp_y$,
$\phi_2=zp_x-xp_z$ and $\phi_3=xp_y-yp_x$ exhibit the $SO(3)$
algebra. Such systems require cubic terms in Lagrangian and/or
Hamiltonian, assuming the primary constraints are linear.
Yang-Mils models fall in this category. More complicated examples
in which constraints of different powers constitute a closed
algebra may be imagined. But such strange systems are not met in
concrete physical models, and it seems that following
sophisticated discussions in this direction does not give us more
insight about gauge theories.

One important point should be added here. It is well known that
\cite{HENOUX}, multiplicative expressions  of constraints (first
or second class) are first class, i.e. their Poisson brackets with
all constraints are at least linear with respect to the
constraints and vanish weakly. It is possible to construct, for
instance, a set of quadratic expressions out of a smaller set of
constraints such that they make a closed non abelian Lie algebra
of Poisson brackets. Assume, for example, four second class
constraints $x,p_x$ and $y,p_y$. One can write ten quadratic
monomials such as $xp_x$,$xy$, $p_xp_y$, etc. Clearly these
constraints are first class and show up a non abelian closed Lie
algebra. We consider a subset of them as follows:
    \be  \begin{array}{ccc}
     \phi_1=x^2 & \phi_2=xp_x & \phi_3=xp_y  \\
      \phi_4=p_x^2 & \phi_5=p_xp_y &  \\
     \phi_6=p_y^2.  &      &\\
   \end{array}  \label{rrr2}\ee
They obey the following closed algebra:
   \be \begin{array}{ccccccc}
   \{,\}   &\phi_1  &\phi_2  &\phi_3  &\phi_4 &\phi_5  &\phi_6  \\
  \phi_1   &  0     &2\phi_1 &0       &4\phi_2&2\phi_3 &0       \\
  \phi_2   &-2\phi_1&0       &-\phi_3 &2\phi_4&\phi_5  &0       \\
  \phi_3   &   0    &\phi_3  &0       &2\phi_5&\phi_6  &0       \\
  \phi_4   &-4\phi_2&-2\phi_4&-2\phi_5&   0   & 0      &0       \\
  \phi_5   &-2\phi_3&- \phi_5&-\phi_6 &    0  &  0     &0       \\
  \phi_6   &0       &   0    &   0    & 0     &  0     &0       \\
   \end{array}  \label{rrr3}\ee
It is clear that the constraints surface described by
$\phi_1\cdots\phi_6$ in (\ref{rrr2}) is the same as given by
$\psi_1\equiv x\approx0$, $\psi_2\equiv p_x\approx0$ and
$\psi_3\equiv p_y\approx0$ which is a mixed system composed of a
pair of second class constraints $(x,p_x)$ together with the first
class constraint $\psi_3$. Such systems are recognized in the
literature \cite{HENOUX, zanelli} as irregular constraints, which
are defined as systems of constraints whose gradients vanish on
the constraint surface. More precisely their Jacobian
$\frac{\partial\phi_a}{\partial \xi_j}$ (where $\xi_j$ are the
phase space coordinates ) is not full rank on the constraint
surface.

If the non abelian algebra of the first class constraints has
originated from the irregular nature of the constraints, then the
most direct way toward an abelian equivalent system is
regularizing the system, i.e. replacing the multiplicative
constraints by the equivalent linear ones. For this reason we
should classify multiplicative constraints into three different
categories:
\begin{enumerate}
    \item Nonlinear constraints of the form $[f(q,p)]^k\approx0$.
    Such a constraint should be linearized, i.e. replaced by linear
    expression $f(q,p)\approx0$. Our evidence for this replacement
    is that a constraint of the form $[f(q,p)]^k\approx0$ has no
meaning other than $f(q,p)\approx0$. This suggestion does not mean
any change in the Hamiltonian or Lagrangian (as in ref.
\cite{zanelli}).
    \item Bifurcating systems of the form $f(q,p)g(q,p)\approx0$.
    In this case the constraint surface obviously "bifurcates" into two different
branches $f(q,p)\approx0$ and $g(q,p)\approx0$. These systems
should be considered as the union of different
    branching $f(q,p)\approx0$ and $g(q,p)\approx0$. Each branch
    should be treated individually. For example, the system given
by $\phi_1=x^2$, $\phi_2=xp_x$ and $\phi_3=xy$ may be regularized
to $\psi_1=x$ (first class), or $\psi_1=x,\psi_2=p_x$ (second
class) or $\psi'_1=y, \psi_2=p_x$ (first class). In each branch
the dynamics of the system is different from the others.
    \item Reducible systems of the form $f(q,p)G(q,p)\approx0$
    where $G(q,p)$ has no finite root. One can replace such a
    constraint with the simpler one $f(q,p)\approx0$. This
    reduction is in fact, the essence of the abelianization
    procedure explained in the  next section.
\end{enumerate}
As the consequence of three kinds of simplifications given above
the number of constraints, as well as their algebra, may change.
In most cases the regularization and reduction procedures
simplifies the algebra of constraints as the above examples show.
However, the more important point is clarifying the first or
second class nature of the system. For example the system given in
(\ref{rrr2}) is not in reality a first class system. Besides
simplifying the algebra, the important point is that the system
given by (\ref{rrr2}) is partly second class (in the $x$, $p_x$
plane). Conclusively, given a set of irregular first class
constraints satisfying a non abelian algebra, there is no
guarantee that the system remains first class after
regularization.

One problem that makes the above analysis more difficult is the
possibility of combining the multiplicative constraints in the
form of more complicated functions. Another problem is that the
simple multiplicative forms may be hidden by changing the
variables describing the phase space or the constraint surface.
Such difficulties may be resolved during the process of
abelianization in which one tries to find the most suitable
coordinates for describing the constraint surface.

As a prescription toward an abelian algebra we propose to
regularize a system of multiplicative constraints in advance. Then
the remaining algebra of first class constraints, if it is still
non abelian, can be abelianized following the method of the next
section. Therefore, in the following we assume that the system is
already regularized. We restrict ourselves to a pure and truly
first class system and try to abelianize the algebra of Poisson
brackets. By truly first class system we mean a system that
remains first class in every description of the constraint
surface.
  \section{Abelianization \label{sec3}}
  Suppose we are given a set of first class constraints
$\phi_i(q,p)$,  $i=1,\cdots m$, satisfying the algebra
   \be \{\phi_i,\phi_j\}=\alpha_{ijk}\phi_k\label{eq1s3}\ee
where in general $\alpha_{ijk}$ depends on $(q,p)$. We wish to
find an equivalent set  of constraints $\widetilde{\phi_r}(q,p)$
where
   \be   \begin{array}{c}
   \phi_i(q,p)=0\Leftrightarrow\widetilde{\phi_r}(q,p)=0  \\
   \{\widetilde{\phi_r}(q,p),\widetilde{\phi_s}(q,p)\}=0, \\
   \end{array} \label{r7e1}\ee
assuming that the system remains first class under the reduction
$\phi_i\rightarrow\widetilde{\phi_r}$. If we are succeeded in this
regard, then it is in principle possible to find a suitable
canonical transformation which transforms all the
$\widetilde{\phi_r}$ to a set of momenta, i.e.
   \be \widetilde{\phi_r}\rightarrow P_r. \label{eq2s3}\ee

In the first step we can choose the first constraint momentum
$P_1$ the same as $\phi_1(q,p)$. It is justified that in principle
there exists a CT which does this task, although it may require
heavy algebraic manipulations. Therefore, suppose under the
desired CT we have
   \be  P_1\equiv\phi_1(q,p).  \label{eq3s3}\ee
Suppose $Q_1$ is the coordinate conjugate to $P_1$ and the other
coordinates change to $\widetilde{q_2},\widetilde{p_2},\cdots$
under the above CT. In practice $Q_1$ should be determined by
solving the differential equation due to $\{Q,P\}=1$ and
$\widetilde{q}$, $\widetilde{p}$ are derived after determining the
corresponding generating function of CT, as we will show in the
example of the next section. In the new coordinates all the
remaining constraints may change, so as to say
   \be  \begin{array}{ccc}
       \phi_1(q,p) & \longrightarrow  & P_1 \\
       \phi_2(q,p) & \longrightarrow &
 \phi_2(Q_1,P_1,\widetilde{q},\widetilde{p}) \\
       \vdots &  \longrightarrow & \vdots \\
     \end{array}   \label{eq4s3}\ee
We first show that $P_1$ should be absent in the remaining
constraints $\phi_i(Q_1,P_1,\widetilde{q},\widetilde{p})$. The
reason is that in general we can expand $\phi_i$ such that:
  \be \phi_i(Q_1,P_1,\widetilde{q},\widetilde{p})= \phi_i(Q_1,0,\widetilde{q},\widetilde{p})
  +P_1f_i(Q_1,P_1,\widetilde{q},\widetilde{p}) \ee
Excluding irregular cases where the gradient of constraints go to
infinity on the constraint surfaces; the Taylor expansion always
makes sense. On the constraint surface we have $P_1=0$. So we have
the equivalency:
  \be  \begin{array}{ccc}
     P_1 &   & P_1 \\
     \phi_2(Q_1,P_1,\widetilde{q},\widetilde{p}) & \longrightarrow &
 \phi_2(Q_1,0,\widetilde{q},\widetilde{p})\equiv\phi_2(Q_1,\widetilde{q},\widetilde{p}) \\
     \vdots &  \longrightarrow & \vdots \\
   \end{array}  \label{rr1}\ee
Now we are at the point to consider the powerful requirement that
the constraints are first class. For example the Poisson brackets
$\{\phi_i(Q_1,\widetilde{q},\widetilde{p}),P_1\}$ vanishes on the
constraint surface described yet in Eq. (\ref{rr1}):
  \be \frac{\partial \phi_i(Q_1,\widetilde{q},\widetilde{p})}{\partial Q_1}=
  \sum_{j=2}^{m}C_{ij}(Q_1,\widetilde{q},\widetilde{p}) \phi_j(Q_1,\widetilde{q},\widetilde{p})
  \;\;\;\ i=2,\cdots m.\label{dife}\ee
Since $P_1$ is absent from LHS of the above equation, no term
proportional to $P_1$ has been written in the RHS. This system of
coupled ordinary linear first order differential equations has an
analytic solution when $C_{ij}$'s are independent of $Q_1$ which
reads in the matrix notation as
  \be \phi(Q_1,\widetilde{q},\widetilde{p})=e^{CQ_1}\phi(0,\widetilde{q},\widetilde{p})  \label{rrr4}\ee
where $C$ is a $(m-1)\times(m-1)$ matrix with elements
$C_{ij}(\widetilde{q},\widetilde{p})$. This is the case for models
in which the structure functions are constant. Since the
exponential part in (\ref{rrr4}) can not vanish for any finite
value of $Q_1$, the necessary and sufficient condition for
vanishing of $\phi_i(Q_1,\widetilde{q},\widetilde{p})$ is
vanishing of $\phi_i(0,\widetilde{q},\widetilde{p})$ which we can
rename them as $\eta_i(\widetilde{q},\widetilde{p})$. So, the
reduction shown in Eq. (\ref{rr1}) goes on one step further to
  \be  \begin{array}{ccc}
     P_1 &  \longrightarrow & P_1 \\  \phi_2(Q_1,P_1,\widetilde{q},\widetilde{p}) & \longrightarrow &
  \phi_2(0,\widetilde{q},\widetilde{p})\equiv\eta_2(\widetilde{q},\widetilde{p}) \\ \vdots &  \longrightarrow & \vdots \\
   \end{array}.  \label{rrs1}\ee
The above case with constant $C_{ij}$ gives the general feature of
the problem. The result is that given $P_1$ as a constraint, the
remaining constraints after redefinition should be independent of
$Q_1$. In other words the $Q_1$-dependent part of them has no root
and can be omitted.

This feature can also be shown for the general case through the
following argument. We come back to the system of linear ODE's
given by Eq. (\ref{dife}). From the theory of differential
equations \cite{corant} it is well-known that the general
solutions should contain $m-1$ constants, which can be chosen as
the initial values $\phi_i(0,\widetilde{q},\widetilde{p})$. In
fact, it can be shown that the general solution
$\phi_i(Q_1,\widetilde{q},\widetilde{p})$ is linear with respect
to the initial values $\phi_i(0,\widetilde{q},\widetilde{p})$.
Just to remind the reader, the proof can be achieved by dividing
the interval $(0,Q_1)$ into an infinite number of segments $\delta
Q_1$; then the linear differential equations (\ref{dife}) show
that after $n$ steps we have
    \be    \begin{array}{l}
      \phi_i(n\delta Q_1,\widetilde{q},\widetilde{p})=\\
    (1+C((n-1)\delta Q_1,\widetilde{q},\widetilde{p}))_{ij}
    (1+C((n-2)\delta Q_1,\widetilde{q},\widetilde{p}))_{jk} \\
      \cdots_n
  (1+C(0,\widetilde{q},\widetilde{p}))_{nl}\phi_l(0,\widetilde{q},\widetilde{p}). \\
    \end{array}    \ee
Although the product of parentheses above can not be exhibited by
a simple exponential (as in Eq.(\ref{rrr4})), the linearity of the
final values of $\phi_i(Q_1,\widetilde{q},\widetilde{p})$ with
respect to initial values $\phi_i(0,\widetilde{q},\widetilde{p})$
is established. In this way one can propose the general form of
the solution of Eqs. (\ref{dife}) as
    \be \phi_i(Q_1,\widetilde{q},\widetilde{p})
  =\sum_j\phi_i^{(j)}(Q_1,\widetilde{q},\widetilde{p})\eta_j(\widetilde{q},\widetilde{p}) \label{gensol}\ee
    where
    \be  \eta_j(\widetilde{q},\widetilde{p})\equiv\phi_j(0,\widetilde{q},\widetilde{p}).  \label{taerif}\ee
The functions $\phi_i^{(j)}(Q_1,\widetilde{q},\widetilde{p})$  are
special solutions of Eqs. (\ref{dife}) with the initial conditions
     \be \phi_i^{(j)}(0,\widetilde{q},\widetilde{p})=\delta_{i}^{j}
      \;\;\;\;\; i,j=2,\cdots m-1.\label{initialvalue}\ee

The general solution (\ref{gensol}) can be viewed as an expansion
in terms of special solution $\phi_i^{(j)}
(Q_1,\widetilde{q},\widetilde{p})$ of the ODE's (\ref{dife}). From
this point of view the functions
$\eta_j(\widetilde{q},\widetilde{p})$ can be interpreted as the
constant (with respect to $Q_1$) coefficients of expansion. It
should be noticed that we did not solve practically the
differential equations (\ref{dife}) which are resulted from the
algebra of the constraints. The point is that the constraints
$\phi_2,\cdots , \phi_m$ in terms of coordinates
$(Q_1,P_1,\widetilde{q},\widetilde{p})$ automatically should
appear in the form of solutions of Eqs. (\ref{dife}). This point
will be seen clearly in the example given in the next section.

Now we claim that the constraint surface given by
$\phi_i(Q_1,\widetilde{q},\widetilde{p})$ is the same as one given
by $\eta_i(\widetilde{q},\widetilde{p})$. It is obvious from
(\ref{gensol}) that $\eta_i(\widetilde{q}, \widetilde{p})=0$ give
rise to $\phi_i(Q_1,\widetilde{q},\widetilde{p})=0$. What about
the inverse deduction? Our assertion here is that if in some
direction the constraint surface can not be described unless some
definite function of $Q_1$ say
$f(Q_1,\widetilde{q},\widetilde{p})$ vanishes, then the equation
$f(Q_1, \widetilde{q},\widetilde{p})=0$ can in principle be solved
to give $Q_1$ as  $Q_1=g(\widetilde{q}, \widetilde{p})$. Then
$P_1$ and $Q_1-g(\widetilde{q},\widetilde{p})$ constitute a second
class constrained system, which we have excluded it from our
consideration. Hence, the necessary and sufficient condition for
$\phi_i(Q_1,\widetilde{q},\widetilde{p})=0$ is
$\eta_i(\widetilde{q},\widetilde{p}) =0$. So we come to a
noticeable result that the redefinition procedure has brought us
to the set of equivalent constraints $P_1$,
$\eta_2(\widetilde{q},\widetilde{p})$, $\cdots$,
$\eta_m(\widetilde{q},\widetilde{p})$, with the property that
$P_1$ commutes with all other constraints. By this procedure we
have decoupled the constraint $\phi_1=P_1$ from others. Decoupling
is done by purging other constraints from canonical conjugate pair
$(Q_1,P_1)$.

Now we can restrict our attention to constraints
$\eta_2,\cdots\eta_m$ which are defined in a smaller phase space
$(\widetilde{q},\widetilde{p})$ where the canonical pair
$(Q_1,P_1)$ are no longer present. Any canonical transformation in
the $(\widetilde{q},\widetilde{p})$ subspace does not affect the
subspace $(Q_1,P_1)$. Therefore, one can in principle repeat the
same procedure once more and in this time assumes that
$\eta_2(\widetilde{q},\widetilde{p})$ is the momentum $P_2$ in
some suitable coordinates. In this way after several stages all
the constraints would be reduced to a set of momenta.

Note should be added that the number of constraints may be
changed, in fact reduced, at any stage of the above process of
abelianization. The reason is that, for example in the  first
stage, linear independence of the constraints
$\phi_i(Q_1,\widetilde{q},\widetilde{p})$, does not necessarily
require that $\phi_i(0,\widetilde{q},\widetilde{p})$ are linearly
independent. Hence, from Eq. (\ref{taerif}) the number of
independent $\eta_i(\widetilde{q},\widetilde{p})$ may be less than
$\phi_i(Q_1,\widetilde{q},\widetilde{p})$.

As a concrete example consider two first class non abelian
constraints $\phi_1$ and $\phi_2$ with the algebra
  \be \{\phi_1,\phi_2\}=\alpha\phi_1+\beta\phi_2. \label{eq2}\ee
By a canonical transformation we map the constraint $\phi_1$ to
momentum $P_1$. After projection $\phi_2$ on the surface $P_1=0$,
the algebra (\ref{eq2}) turns to:
  \be \{P_1,\phi_2\}=\beta\phi_2(Q_1,\widetilde{q},\widetilde{p})\label{eq3}\ee
The constraint $\phi_2$ can be found from the  differential
equation
  \be \frac{\partial\phi_2}{\partial Q_1}=-\beta\phi_2 \label{eq4} \ee
as
  \be \phi_2(Q_1,\widetilde{q},\widetilde{p})=
  \eta(\widetilde{q},\widetilde{p})\exp{(-\int_0^{Q_1}\beta dQ'_1)}
 \label{eq5}\ee
where
$\eta(\widetilde{q},\widetilde{p})=\phi_2(0,\widetilde{q},\widetilde{p})$.
This is a realization of the solutions given in Eq. (\ref{gensol})
for the general case. Since the exponential function does not
vanish for finite values of its argument, vanishing of $\phi_2$
could be only due to $\eta(\widetilde{q},\widetilde{p})$. The
constraints $\phi_1$ and $\phi_2$ are equivalent to $P_1$ and
$\eta(\widetilde{q},\widetilde{p})$, where
  \be \{P_1,\eta\}=0 \label{eq6}\ee
Then, we can make a CT to canonical variables in which
$\eta(\widetilde{q},\widetilde{p})$ is $P_2$. One may wonder if
$\beta(Q_1,q,p)$ is such that $\int\beta dQ_1=\ln {f(Q_1,q,p)}$
and the equation $f(Q_1,q,p)=0$ has some finite roots, then one
may no longer exclude vanishing of the exponential part in
(\ref{eq5}). If this is the case, the equivalent constraints are
$P_1$ and $f(Q_1,q,p)$ where $f(Q_1,q,p)$ can be solved for $Q_1$.
This leads to a second class system which has been excluded
before.
\section{Examples on $SO(3)$ and $SO(4)$ algebras \label{so13}}
In this section we apply our method to some examples. A famous non
abelian algebra is the angular momentum algebra in three
dimensional configuration space. Let us at first stage show how a
$SO(3)$ algebra of constraints may emerge. We may assume that for
a rotational invariant Hamiltonian, $L_x$, $L_y$ and $L_z$ are
given as primary constraints. Consistency condition of primary
constraints then gives no further secondary constraints and the
set of constraints terminate here.

It is also possible to consider more realistic examples in which
the angular momentum algebra emerge in a natural way. For example
the Lagrangian
 \be L=\frac{1}{2}\dot{{\bf X}}^2-V({\bf X}^2)-{\bf\xi}.{\bf L} \label{ex2} \ee
where ${\bf X}\equiv(x,y,z)$ and ${\bf \xi}=(\xi_x,\xi_y,\xi_z)$
constitute a six dimensional configuration space, in which
$L_i=\epsilon_{ijk}x_j\dot{x_k}$. In phase space $\pi_x$, $\pi_y$
and $\pi_z$, the momenta conjugate to $\xi_x$, $\xi_y$ and
$\xi_z$, are primary constraints and the total Hamiltonian reads
 \be H_T=\frac{1}{2}{\bf P}^2+V({\bf X}^2)+{\bf\xi}.{\bf L}
 +{\bf\lambda}. {\bf \pi} \label{ex3} \ee
where ${\bf P}\equiv(P_x,P_y,P_z)$ represents the momenta
conjugate to ${\bf X}$,
${\bf\lambda}\equiv(\lambda_x,\lambda_y,\lambda_z)$ shows Lagrange
multipliers and $L_i=\epsilon_{ijk}x_jp_k$. The consistency
conditions of primary constraints $\pi_i$ give secondary
constraints $L_i$ and no further constraint emerges from the
consistency of $L_i$. First level constraints are abelian while
the second level constraints $L_i$ obey the non abelian $SO(3)$
algebra with constant structure functions $\epsilon_{ijk}$, i.e.
 \be \{L_i,L_j\}=\epsilon_{ijk}L_k \;\;\;\;\;\ i,j=1, 2, 3. \label{ex1}\ee
It is also possible to get the $SO(3)$ algebra from the Lagrangian
 \be L=\frac{1}{2}{\bf\dot{X}}^2-V({\bf X}^2)-e^wL_x-L_y \label{ex4} \ee
where $w$ is a variable. Consistency of $p_w$ gives $L_x$, $L_y$
and $L_z$ respectively as the second, third and forth level
constraints.

The above expressions are related to a one particle system with
rotational invariance. Similar treatments may be considered for
multi-particle systems. Now let us examine our method of
abelianization to different examples of this character.

\subsection{One particle $SO(3)$ model}
Now let us go through the abelianization procedure of the $SO(3)$
algebra of constraints. As stated in the previous section we
should first find a CT that transforms for example $L_1=xp_y-yp_x$
to a momentum $P_1$. The conjugate coordinate $Q_1$ should be
determined such that
 \be \{Q_1,P_1\}=1. \label{ex5}\ee
A possible solution for $Q_1$ is
 \be Q_1=\tan^{-1}(\frac{y}{x}). \label{ex7}\ee
As is apparent, $P_1$ and $Q_1$ are functions of subspace
$(x,y;p_x,p_y)$. Hence, we can exclude the subspace $(z,p_z)$.
Reminding the standard method \cite{goldestein} for extracting a
CT from a generating function, the following generator can be used
 \be F(x,y,P_1,P_2)=P_1\tan^{-1}(\frac{y}{x})+P_2f(x,y). \label{ex8}\ee
Transformation relations then gives $Q_1$ as in (\ref{ex7}) and
$Q_2=f(x,y)$. Imposing the task $xp_y-yp_x=P_1$ on the relations
$p_x=\frac{\partial F}{\partial x}$ and $p_y=\frac{\partial
F}{\partial y}$ also gives $x\frac{\partial f}{\partial
x}-y\frac{\partial f}{\partial y}=0$. In this way the canonical
pair $(Q_2,P_2)$ can be given as
 \be \begin{array}{c}
        Q_2=\frac{1}{2}\ln(x^2+y^2) \\
        P_2=xp_x+yp_y. \\
      \end{array} \label{ex9}\ee
Renaming the variables $(Q_1,P_1;Q_2,P_2)$ as
$(\varphi,p_{\varphi};\psi,p_{\psi})$ the old variables can be
written in terms of the new ones as
 \be \begin{array}{ll}
 z=Z & p_z=P_Z \\
 x=e^{\psi}\cos{\varphi}  & p_x=e^{-\psi}(p_{\psi}\cos{\varphi}-p_{\varphi}\sin{\varphi})
 \\
  y= e^{\psi}\sin{\varphi} & p_y=e^{-\psi}(p_{\varphi}\cos{\varphi} +p_{\psi}\sin{\varphi}). \\
 \end{array} \label{ex10}\ee
In terms of the new variables the constraints $(L_1,L_2,L_3)$ are:
 \be   \begin{array}{ll}
 L_1=p_{\varphi} \\
    L_2=\eta_2(z,p_z,\psi,p_{\psi})\cos\varphi-\eta_3(z,p_z,\psi,p_{\psi})\sin\varphi \\
  L_3=\eta_2(z,p_z,\psi,p_{\psi})\sin\varphi+\eta_3(z,p_z,\psi,p_{\psi})\cos\varphi \\
 \end{array}   \label{ex11}\ee
where
 \be \begin{array}{ll} \eta_2(z,p_z;\psi, p_{\psi},p_{\varphi})\equiv -zp_{\varphi} e^{-\psi}, \\
    \eta_3(z,p_z;\psi, p_{\psi})\equiv e^{-\psi}zp_{\psi}-e^{\psi}p_z .\end{array}\label{ex12}\ee
From the angular momentum algebra we have
 \be   \frac{\partial L_2}{\partial\varphi}=-L_3, \;\;\;\;\;\
 \frac{\partial L_3}{\partial\varphi}=L_2.
 \label{ex14} \ee
These are the same differential equations as (\ref{dife}). Eqs.
(\ref{ex11}) are in fact the solutions of Eqs. (\ref{ex14}) with
respect to the variable $Q_1=\varphi$. As is seen, Eqs.
(\ref{ex11}) are in the form given in Eqs. (\ref{gensol}). The
solution $L_2\propto -\sin\varphi$ and $L_3\propto\cos\varphi$ is
the one with initial condition $L_2(\varphi=0)=0$ and
$L_3(\varphi=0)=1$, as stated in Eq. (\ref{initialvalue}), and the
solution $L_2\propto\cos\varphi$ and $L_3\propto\sin\varphi$
satisfy $L_2(\varphi=0)=1$ and $L_3(\varphi=0)=0$. Since
$\eta_2\approx0$ on the surface $p_{\varphi}=0$ we see that the
set of constraints $(L_1,L_3)$ finally reduces to $p_{\varphi}$
and $\eta_3$ which commute with each other.

Important notice should be added that our transformation here is
not acceptable globally. In fact, at $x=y=0$ the transformation is
singular. Therefore as indicated in some references \cite{gomis}
the abelianization process of $SO(3)$ algebra can be done just
locally. We remind the reader that far from the origin the
constraint surface given by functions $L_1$, $L_2$ and $L_3$ is
the same as given by two of them. In fact, since ${\bf x}.{\bf
L}={\bf p}.{\bf L}=0$ the constraints $L_1$, $L_2$ and $L_3$ are
reducible, provided that ${\bf x}\neq 0$ and/or ${\bf p}\neq 0$.
It seems that this subtle point is the essence that the reference
\cite{L3} has not given a clear statement that the $SO(3)$ gauge
system is abelianizable or not. However, an expanded version of
$SO(3)$ gauge system is shown to be abelianizable in \cite{L4}.
\subsection{Two particle $SO(3)$ model}
We can extend the above model to a system of two particles with
coordinates $(x,y,z)$ and $(a,b,c)$. The angular momentum
components of this system read:
  \be \begin{array}{c}
        L_1=xp_y-yp_x+ap_b-bp_a \\
        L_2=yp_z-zp_y+bp_c-cp_b \\
        L_3=zp_x-xp_z+cp_a-ap_c \\
      \end{array}  \label{pang}\ee
with the same algebra as before.  At the first stage we consider
$L_1$  as (twice) the momentum $p_{\varphi}$ in a new coordinate
system. Using the results of one particle system it is easily seen
that $\varphi=\varphi_1+\varphi_2$  is the coordinate conjugate to
$p_{\varphi}$ where
  \be \begin{array}{cc}
   \varphi_1=\arctan(\frac{y}{x}) & \varphi_2=\arctan(\frac{b}{a}) \\
      \end{array}  \label{new2pco} \ee
In this way we should consider
 \be \begin{array}{cc}
 \varphi=\varphi_1+\varphi_2 & \chi=\varphi_1-\varphi_2\\
  p_{\varphi}=\frac{1}{2}(p_{\varphi_1}+p_{\varphi_2}) & p_{\chi}=\frac{1}{2}(p_{\varphi_1}-p_{\varphi_2}) \\
     \end{array}  \label{def2p} \ee
as new coordinates, where
 \be p_{\varphi_1}=xp_y-yp_x \;\;\;\;\ p_{\varphi_2}=ap_b-bp_a \label{phis}\ee
The canonical transformation which gives the above new coordinates
act in the four dimensional space of $(x,y;a.b)$  and the
corresponding momenta. This transformation should be accompanied
by the following new variables
 \be \begin{array}{cc} \psi_1=\frac{1}{2}\ln(x^2+y^2) &
 \psi_2=\frac{1}{2}\ln(a^2+b^2) \\ p_{\psi_1}=xp_x+yp_y &
 p_{\psi_2}=ap_a+bp_b \end{array} \label{rs} \ee
Inverting equations (\ref{new2pco}), (\ref{def2p}), (\ref{phis})
and (\ref{rs}) gives old variables in terms of new ones. Inserting
them into (\ref{pang}) gives $L_2$ and $L_3$ projected on the
subspace $p_{\varphi}=0$ as:
 \be \begin{array}{l}
 L_2=-p_zr_1\sin\frac{\varphi+\chi}{2}+\frac{z}{r_1}(p_{\chi_1}\cos\frac{\varphi+\chi}{2}+p_{r_1}\sin\frac{\varphi+\chi}{2})\\
  -\frac{1}{r_2}(cp_{\chi}\cos\frac{\varphi-\chi}{2}-(r_2^2p_c-cp_{r_2})\sin\frac{\varphi-\chi}{2}) \\
 L_3=p_cr_2\cos\frac{\varphi-\chi}{2}-\frac{c}{r_2}(p_{\chi}\sin\frac{\varphi-\chi}{2}+p_{r_2}\cos\frac{\varphi-\chi}{2})\\
-\frac{1}{r_1}(zp_{\chi}\sin\frac{\varphi+\chi}{2}+(r_1^2p_z-zp_{r_1})\cos\frac{\varphi+\chi}{2}).\\
     \end{array} \ee
These can be rewritten as
 \be \begin{array}{c}
   L_2=-M_2\sin\frac{\varphi}{2}+M_3\cos\frac{\varphi}{2} \\
   L_3=M_3\sin\frac{\varphi}{2}+M_2\cos\frac{\varphi}{2} \end{array} \ee
where
 \be \begin{array}{c}
       M_2=(r_2p_c+r_1p_z-\frac{zp_{r_1}}{r_1}-\frac{cp_{r_2}}{r_2})\cos\frac{\chi}{2} \\
       +(\frac{z}{r_1}+\frac{c}{r_2})p_{\chi}\sin\frac{\chi}{2}\\
       M_3=(r_2p_c-r_1p_z+\frac{zp_{r_1}}{r_1}-\frac{cp_{r_2}}{r_2})\sin\frac{\chi}{2}  \\
       +(\frac{z}{r_1}-\frac{c}{r_2})p_{\chi}\cos\frac{\chi}{2}\\
     \end{array} \ee

 As in (\ref{ex14}) we see that $L^{(2)}_2=\cos\frac{\varphi}{2} ,L^{(2)}_3=\sin \frac{\varphi}{2}$ and
 $L^{(3)}_2=-\sin\frac{\varphi}{2}, L^{(3)}_3=\cos\frac{\varphi}{2}$ are special solutions of angular
 momentum algebra with initial condition (\ref{initialvalue})
 provided that $L_1=2p_{\varphi}$. As discussed in the previous
 section the constraints $L_2$ and $L_3$ reduce to $M_2$ and $M_3$
 in order that the system remain first class. Hence, the constraint
 $p_{\varphi}$ commutes with $M_2$ and $M_3$  and decouples from the
 algebra of first class constraints.

 Now it seems that we should repeat our procedure to abelianize the remaining two constraints. However,
  a little care shows that we can combine $M_2$ and $M_3$ to get
 \be \begin{array}{c}
 S_2=(r_2p_c-r_1p_z+\frac{zp_{r_1}}{r_1}-\frac{cp_{r_2}}{r_2}) M_2-(\frac{z}{r_1}+\frac{c}{r_2})p_{\chi}M_3 \\
 S_3=(r_2p_c+r_1p_z-\frac{zp_{r_1}}{r_1}-\frac{cp_{r_2}}{r_2})M_3-(\frac{z}{r_1}-\frac{c}{r_2})p_{\chi}M_2  \\
     \end{array} \ee
 so that
 \be S_2=F\cos\frac{\chi}{2} \;\;\;\;\ S_3=F\sin\frac{\chi}{2} \ee
 where
 \be \begin{array}{c}
 F=(r_2^2p_c^2-r_1^2p_z^2)+(\frac{c^2(p_{\chi}^2+p_{r_2}^2)}{r_2^2}-\frac{z^2(p_{\chi}^2+p_{r_1}^2)}{r_1^2}) \\
  -2(cp_cp_{r_2}-zp_zp_{r_1}) \end{array}  \ee
 If one demands either $\sin\frac{\chi}{2}=0$ or
 $\cos\frac{\chi}{2}=0$ to satisfy one of the constraints, then the
 condition $F=0$ should also be imposed to satisfy the other
 constraint, which leads to a seconds class system. Therefore the
 only way is to consider just $F=0$ as the only remaining
 constraint. In this way the system is reduced to two constraints
 $p_\varphi$ and $F$, which are apparently abelian.

\subsection{One particle $SO(4)$ model }
  As a third example consider a constrained system with $SO(4)$ gauge symmetry with
  \be L_{ij}=x_ip_j-x_jp_i, \;\;\;\;\ i,j=1,2,3,4\ee
  The algebra of Poisson brackets of constraints read:
   \be  \{L_{ij},L_{kl}\}=\delta_{ik}L_{jl}-\delta_{il}L_{jk}+\delta_{jl}L_{ik}-\delta_{jk}L_{il}  \ee
  Clearly constraints with no common indexes commute. For example
  $\{L_{12},L_{34}\}=0 $. Assume simultaneous change of variables
  in $(1-2)$ as well as $(3-4)$ surfaces similar to $SO(3)$ case, as follows:
  \be \begin{array}{ll}
  \phi=\arctan(\frac{x_2}{x_1}) & p_{\phi}=x_1p_2-x_2p_1 \\
  \sigma=\frac{1}{2}\ln(x_1^2+x_2^2) & p_{\sigma}=x_1p_1+x_2p_2 \\
  \theta=\arctan(\frac{x_4}{x_3}) & p_{\theta}=x_3p_4-x_4p_3 \\
  \rho=\frac{1}{2}\ln (x_3^2+x_4^2)& p_{\rho}=x_3p_3+x_4p_4 \\
  \end{array} \label{so41}\ee
  Inverting the above relations and rewriting the constraints
  in terms of the new variables, we get:
  \be \begin{array}{ll}
  L_{12}=p_{\varphi} & L_{23}=F\cos\theta\sin\varphi \\
  L_{13}=F\cos\theta\cos\varphi & L_{24}=F\sin\theta\sin\varphi \\
  L_{14}=F\sin\theta\cos\varphi & L_{34}=p_{\theta} \\      \end{array} \ee
  where we have reduced the expressions on the surface $p_{\varphi}=p_{\theta}=0$ and
  \be F=e^{\sigma-\rho}p_{\sigma}-e^{\rho-\sigma}p_{\rho}.  \ee
  A little care shows that the $\sin$`s and $\cos$`s in
  (\ref{so41}) provide special solutions of the differential equations due to
  the algebra of constraints (in terms of new variables). In this way the non-abelian system
  of 6 constraints $L_{ij}$ reduced to a system of 3 abelian constraints $p_{\varphi}, p_{\theta},F$.
\section{Concluding remarks \label{conclusion}}
Our main objective in this paper is that the algebra of Poisson
brackets of first class systems may be simplified considerably. In
fact, a complicated algebra may be escaped during the procedure of
determining the constraints from the very beginning. This
possibility has not been discussed here. The other possibility is
simplifying the given constraints (regardless how they are
produced) by using legal algebraic methods. The best thing that
one may desire is an abelian system in which all of the
constraints commute. For this reason one should try to find the
most suitable coordinates describing either the whole phase space
or the constraint surface.

One may use canonical transformations to change the coordinates of
phase space. This process does not change the algebra of
constraints. However, it may be used, for example, to clarify the
multiplicative nature of the constraints. On the other hand, the
constraint surface may be demonstrated by vanishing different sets
of functions, which may have different algebras of Poisson
brackets. This possibility, i.e. "redefinition" of constraint
surface, motivates our desire of abelianization of the constraint
systems.

In this paper we first observed that ordinary quadratic
Lagrangians lead to abelian constraints which are linear with
respect to phase space coordinates. Non abelian algebras require
at least cubic or more complicated functions of coordinates and
velocities in the Lagrangian.

We also showed with some examples that non abelian systems may
appear because of unsuitable choice of constraints. In other
words, one may solve the constraint equations to find the most
simple way to describe the constraint surface. This procedure is
in fact redefinition of the constraint surface. The main reason
behind these calculations is to find the most suitable coordinates
in which the constraint surface is described by vanishing or
constancy of some phase space coordinates.

Although we were not able to prove a wide-standing theorem, we
observed that sometimes the non abelian constraints (in suitable
coordinates) contain factors, such as exponentials, that have no
root in the finite region of the range of variables. Therefore,
one may redefine the constraints by omitting such factors. This
procedure is formulated in a systematic way by solving the
differential equations due to the algebra of first class
constraints. This is in fact the essence of our method of
abelianization of the constraints.

The method may involve heavy algebraic manipulations, even in the
seemingly simple step of canonical transformations. However, one
can use it in any concrete example of constrained systems, despite
the fact that for more complicated systems, like any other
mathematical method, lengthy calculations are needed to get the
final result. In other words it seems that all of the constrained
systems may in principal be abelianized and no description can be
found about a system which is "intrinsically non-abelianizable".

We showed that our method works well for a one-particle as well as
two-particle system with $SO(3)$ symmetry. The problem is also
solved for $SO(4)$ symmetry and one may find it straightforward to
follow the same procedure for $SO(5)$ and more generally for
$SO(N)$. Other simple Lie algebras may also be considered in the
same way.

There exist, however, two important points which should be
considered carefully, i.e. "regularity" and "locality". The first
point is the case when we encounter multiplicative constraints. We
suggest that multiplicative constraints should be regularized at
each step of calculation. This means that they should be replaced
by simple roots of the corresponding expressions.

One important feature in this regard is that during simplifying
the multiplicative constraints, the first class nature of the
system may be altered. In general, there is no guarantee that the
system remains first class after such simplification of the
constraints. This point makes us to take the necessity of
regularization of the constraints more serious, since otherwise it
is possible to have a first class algebra of constraints out of a
number of second class ones.

The next important point is that most of the above mathematical
manipulations (i.e. regularization, canonical transformation and
redefinition of constraints) which we use to find a simple
description of constraint surface are valid locally, and may fail
for some singular points or finite regions of the constraint
surface. For example, in the case of $SO(3)$ algebra we showed
that the system can be abelianized everywhere except the origin of
the phase space. In this way we can divide first class systems
into "globally abelianizable" and "locally abelianizable".

We mention briefly that both the above problems are not
difficulties with respect to our method of abelianization. They
may be considered as intrinsic features of the constrained system
itself, as it is described from the very beginning. For example,
the problem of locality arises whenever the algebra of constraints
is not valid globally throughout the whole phase space. Therefore,
it is obvious that any method may encounter difficulties in the
regions where the algebra is different.

The question may arise, however, that is there any advantage in
employing the abelianized system instead of the original one? For
instance what is preference of using the abelian constraints
$p_{\varphi}$ and $\eta$ in the case of $SO(3)$, instead $L_1$,
$L_2$ and $L_3$? The answer depends on different applications of
the first class constraints in physical problems such as gauge
symmetry, quantization procedure, counting the physical degrees of
freedom and so on. We postpone such analysis to future works. We
reserve this possibility that maybe in some cases it is better to
keep a non abelian and reducible algebra of constraints instead of
change it into a non reducible and abelian one. The reason may be
the better possibility of tracking the physical symmetries such as
rotation.

\textbf{Acknowledgements:} The authors would like to thanks
Institute for Research in Fundamental Sciences (IPM) for financial
support.

\end{document}